\documentclass[aps,prb,twocolumn,superscriptaddress,floatfix,longbibliography]{revtex4-2}

\usepackage[utf8]{inputenc}
\usepackage{graphicx}
\usepackage{braket}
\usepackage{bm}
\usepackage{upgreek}
\usepackage{mathtools}
\usepackage{array}
\usepackage{physics}

\usepackage{hyperref}

\begin{document}
\title{Wavefunctions and oscillator strengths of Rydberg excitons in cuprous oxide quantum wells}

\author{Leon Kühner}
\author{Patric Rommel}
\author{Jörg Main}
\email[Email: ]{main@itp1.uni-stuttgart.de}
\affiliation{Institut für Theoretische Physik I, Universität
  Stuttgart, 70550 Stuttgart, Germany}
\author{Stefan Scheel}
\author{Pavel A. Belov}
\affiliation{Institut für Physik, Universität Rostock,
  Albert-Einstein-Straße 23-24, 18059 Rostock, Germany}

\date{\today}

\begin{abstract}
We investigate the eigenstates, that is, the wavefunctions of Rydberg
excitons in cuprous oxide quantum wells and derive expressions relating
them to the oscillator strengths of different exciton states.  Using
the B-spline expansion, we compute the wavefunctions in coordinate
space and estimate the oscillator strengths.  The symmetry properties
of the states and the non-separability of the wavefunctions are
illustrated.  Wavefunctions associated with resonances above the
scattering threshold, in particular those of bound states in the
continuum as well as their partner states, are also given.
\end{abstract}

\maketitle

\section{Introduction}
\label{sec:intro}
The interaction of the electromagnetic field with matter forms the
foundation of quantum optics and provides practical applications in
material science.
Investigating the spectral features originating from the light-matter
interaction provides access to the electronic and optical properties
of direct bandgap semiconductors.
In these spectra, the single-particle response in the vicinity of the
absorption edge is complicated by nonlinearities resulting from
many-particle correlations, such as two-particle correlations of the
charge carriers, electrons and holes, forming an exciton.
The importance of excitons in optical transitions has been recognized
since their first experimental observation in bulk cuprous
oxide~\cite{Hayashi1950,Gross1,Gross2}.

Excitons are characterized on the one hand by their binding energy
which specifies how strongly the Coulomb interaction of the charge
carriers dominates over their kinetic motion~\cite{klingshirn2007semiconductor}.
On the other hand, the strength of exciton-light coupling is measured
by the radiative decay rate, or the oscillator strength.
The oscillator strengths of excitons in bulk semiconductors have been
theoretically studied since the early works in the
1960's~\cite{Elliott1957,Toyozawa1958,Hermanson1966}.
Although the lower-lying exciton states in bulk cuprous oxide have
been investigated early on
\cite{Elliott1961,Rustagi1973,Denisov1973,Agekyan1974,Rashba}, the
ratios of the oscillator strengths for the entire Rydberg series of
exciton states have only been recently described in detail in
Ref.~\cite{Schweiner17c}.
It turns out that the two-particle correlations induced by the Coulomb
interaction significantly alter the wavefunctions of the charge carriers.
The exciton wavefunction differs from just a simple overlap of the
wavefunctions of single carriers.
Indeed, the optical matrix element is defined by the derivative of the
exciton wavefunction at zero separation of electron and hole, i.e., at
the point where the carriers recombine.

For confined excitons the optical transitions were initially measured
and reported in Refs.~\cite{Shinada1966,Chemla1984}.
The theory of excitons and their radiative characteristics in
GaAs-based quantum wells (QWs) is well
developed~\cite{Ekenberg1987,ivchenko,Belov2019}.
The spectral peaks of several lowest exciton resonances in
high-quality GaAs/AlGaAs structures are rather well described by the
excitonic contribution to the dielectric
polarization~\cite{Andreani1991,Voronov2007,Khramtsov2016}.
However, for Rydberg excitons confined in cuprous oxide QWs, only
their energetic properties have been well understood
recently~\cite{Scheuler2024,Belov2024}.
The effect of the quantum confinement depends on the relative sizes of
the particular exciton states and the confining QW.
For a given QW width, the ground state of the size of the Bohr radius
can be in a weak confinement regime, whereas a highly-excited Rydberg
state can be strongly confined.
In the former case, the Coulomb interaction dominates and the
confinement can be considered as a small perturbation.
In the latter case, the barriers squeeze the exciton wavefunction, and
thus distort its spherical symmetry and significantly modify the
energy spectrum and the selection rules.
Such a squeezing also affects the behavior of the wavefunction at the
origin, hence it changes the radiative properties.

Optical properties of Rydberg excitons in cuprous oxide nanostructures
had previously been analytically studied in
Refs.~\cite{Ziemkiewicz2020,Ziemkiewicz2021a,Ziemkiewicz2021b,Ziemkiewicz2021c}.
To obtain their analytical results on the excitonic susceptibility and
the absorption, several approximations such as a simplified 2D Coulomb
potential~\cite{Duclos2010} depending only on the radial coordinate in
the QW plane and allowing an exact separation of the variables in the
initial Hamiltonian, had to be employed in these works.
Numerical approaches based on the solution of the 2D Schr\"{o}dinger
equation were developed for calculation of the absorption of excitons
in TMDC-based heterostructures and in phosphorene
monolayers~\cite{Brunetti2018,Brunetti2019}.
Here, we extend these results by taking into account the full 3D
Coulomb potential and discuss the effects of the symmetry properties
of the exciton wavefunction on optical transitions.

It is well known that the linewidth broadening of the exciton
transitions results from contributions from both radiative and
nonradiative processes. The nonradiative broadening includes scattering processes, viz.\
electron-hole and electron-impurity scattering, and the interaction
with phonons~\cite{Fr1954,Klaehn1976,Kaz14}.
A rigorous theory of the interaction of electrons with phonons based
on first-principle considerations is reviewed in
Ref.~\cite{Giustino2017} and generalized to exciton-phonon coupling in
Ref.~\cite{Antonius2022}.
The scattering off phonons can be suppressed at low temperature.
The effects from imperfections of the structure are diminished for
high-quality samples.
Moreover, we have shown recently~\cite{Aslanidis2025} that for
particular strengths of the quantum confinement, the nonradiative
electron-hole scattering can be damped and so-called bound states in
the continuum (BICs)~\cite{Wigner1929,Stillinger1975,Hsu2016,Schiller2024,Happ2025,Valero2025}
of the exciton states can appear.
They are caused by the destructive interference of two adjacent
resonances, associated with different subbands of the same
symmetry~\cite{Fri85b}.
As a result, the linewidth broadenings of electron-hole BICs are
mainly defined by the radiative exciton properties.

In this article, we investigate the eigenstates, i.e., the
wavefunctions of excitons in cuprous oxide QWs within a hydrogenlike
model and derive a relationship between the oscillator strengths of
different exciton states.
The description of the radiative characteristics of excitons in
cuprous oxide is more complicated than in GaAs-based semiconductors.
For Cu$_{2}$O, the momentum matrix element between the conduction 
and valence bands $p_{cv}$ vanishes due to same parity of the
bands~\cite{Elliott1957}.
Therefore, one has to analyze the symmetries of operators in the
next-to-leading order terms in the momentum.
Moreover, in contrast to bulk excitons~\cite{Schweiner17c}, due to the
QW confinement, the total wave vector is no longer a good quantum
number, but the projection of the wave vector into the QW plane is
still conserved.
As a result, the oscillator strength involves the derivative of the
exciton wavefunction over the in-plane coordinate at the point at
which the carriers recombine.
Numerical calculations of the wavefunctions of different exciton
states allow us to obtain the relative oscillator strengths of exciton
transitions as a function of the QW width.
The non-separability of the wavefunctions as well as their symmetry
properties~\cite{Lax2001} governing the oscillator strengths are
deduced.

Our method for the numerical evaluation of the Schrödinger equation is
based on the expansion of the exciton wavefunction over a B-spline
basis and described in detail in Refs.~\cite{Scheuler2024,Belov2024}.
Particular attention is paid to the wavefunctions of the Rydberg
excitons and electron-hole resonances in the continuum of lower
quantum-confinement subbands as the latter generate the BICs.
The radiative properties of BICs, in comparison with their partner
states, are also investigated.
We note that, besides the oscillator strength, the numerically
obtained wavefunctions can be used for the calculation of other
physical observables such as the energy or the average radius,
or of effects like the Stark shift via perturbative methods~\cite{Landau}.

The article is organized as follows.
In Sec.~\ref{subsec:ExcitonsQW} we set up the general theoretical
framework for describing Rydberg excitons in a quantum well structure.
We will specify the particular Cu$_{2}$O-based QW and describe
numerical methods used to compute the exciton wavefunctions.
In Sec.~\ref{subsec:oscillatorTheory}, we derive the
relationship between the oscillator strength and the wavefunction,
allowing us to calculate the exciton radiative broadening.
A detailed discussion of the results for bound states, resonant
states, as well as for the BICs is provided in Sec.~\ref{RandD}.
Technical details are relegated to the Appendix.

%%%%%%%%%%%%%%%%%%%%%%%%%%%%%%%%%%%%%%%%%%%%%%%%%%%%%%%%%%%%%%%%%%%%%%%
\section{Theory and methods}
\label{MandM}
\subsection{Excitons in QWs}
\label{subsec:ExcitonsQW}
\begin{figure}
  \includegraphics[width = 0.9\columnwidth]{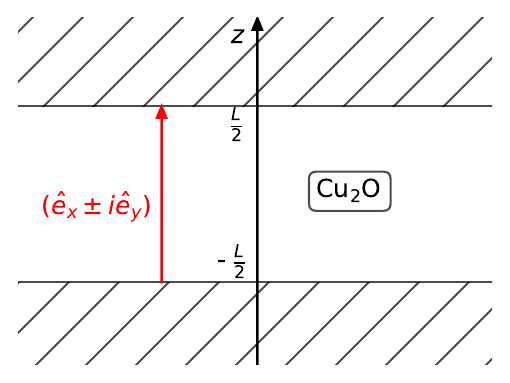}
  \caption{Sketch of the cuprous oxide QW system with width
    $L$ along the $z$ axis. The red arrow indicates a circularly
    polarized incident laser beam producing excitons with angular
    momentum quantum number $m=\pm 1$.}
\label{fig:QW_sketch}
\end{figure}
In this section, we consider an exciton in a low-di\-mensional
structure.
This could be either a heterostructure~\cite{ivchenko} with a thin
film sandwiched by another material, or a finite-sized crystal
surrounded by vacuum or air~\cite{Belov2024}.
For excitons in a thin film, or simply a QW, oriented perpendicular to
the $z$ axis as sketched in Fig.~\ref{fig:QW_sketch}, electron and
hole coordinates are translationally invariant only in the QW plane,
but not along the perpendicular direction.
Within a two-band model, after separating the center-of-mass motion in
the QW plane and introducing the polar coordinates $(\rho,\phi)$ for
relative electron and hole motion in this plane, the Hamiltonian of
the electron-hole pair in the QW structure reads
\begin{align}
	H =   & E_\mathrm{g} - \frac{\hbar^2}{2m_{\mathrm{e}}}\frac{\partial^2}{\partial z_\mathrm{e}^2} - \frac{\hbar^2}{2m_{\mathrm{h}}}\frac{\partial^2}{\partial z_\mathrm{h}^2}-\frac{\hbar^2}{2\mu} \left( \frac{\partial^2}{\partial \rho^2}+ \frac{1}{\rho} \frac{\partial}{\partial \rho} - \frac{m^2}{\rho^2}\right)\nonumber \\ 
	& - \frac{e^2}{4\pi \epsilon_0 \epsilon \sqrt{\rho^2+(z_\mathrm{e}-z_\mathrm{h})^2}} + V_\mathrm{e}(z_\mathrm{e}) +V_\mathrm{h}(z_\mathrm{h}) \, .
\label{eq:Hamiltonian}
\end{align}
Here, the effective masses of the electron $m_{\mathrm{e}}=0.99\, m_{0}$
and the hole $m_{\mathrm{h}}=0.69\, m_{0}$ in the bulk cuprous oxide
semiconductor~\cite{Schweiner16b} define the kinetic terms along the
confinement direction, $E_\mathrm{g} = 2.17208$~eV is the energy gap.

The potentials $V_{\mathrm{e},\mathrm{h}}(z_{\mathrm{e},\mathrm{h}})$
confine the motion of electron and hole across the QW, along the $z$
direction.
These are the potentials which break the spherical symmetry in the
bulk, reducing it to mere translational invariance in the QW plane.
Thus, although in the plane we transform to relative coordinates,
$z_{\mathrm{e}}$ and $z_{\mathrm{h}}$ remain unchanged.
For a finite-sized crystal surrounded by vacuum or air the confinement
potential of the QW can be well approximated by an infinite potential well,
\begin{equation}
\label{eqV}
V_{\mathrm{e,h}}(z_{\mathrm{e,h}}) = \left\{
  \begin{array}{lr}
    0 & \mbox{ if  } |z_{\mathrm{e,h}}|<L/2 \\
    \infty & \mbox{ if  } |z_{\mathrm{e,h}}| \ge L/2
  \end{array}
\right. .
\end{equation}
This potential introduces the electron and hole quantum-confinement
subbands $E_{\mathrm{e}i,\mathrm{h}j}$ where $i$ and $j$ correspond to
independent indices $k$ in the wavefunctions for electron and hole
\begin{equation}
\label{eq:1Dconfinmentfunctions}
\phi_{k}(z_\mathrm{e,h}) = \left\{
  \begin{array}{lr}
    \sqrt{\frac{2}{L}} \cos{\left(\frac{\pi k}{L} z_{\mathrm{e,h}} \right)} & \mbox{ if  } k=1,3,5,\ldots \\
    \sqrt{\frac{2}{L}} \sin{\left(\frac{\pi k}{L} z_{\mathrm{e,h}} \right)} & \mbox{ if  } k=2,4,6,\ldots
  \end{array}
\right. .
\end{equation}
The kinetic energy of the relative electron-hole motion in the QW
plane is governed by the 2D Laplace operator which depends on the
distance $\rho$ between electron and hole in the plane.
In Eq.~\eqref{eq:Hamiltonian}, the parameter 
$\mu=m_{\mathrm{e}}m_{\mathrm{e}}/(m_{\mathrm{e}}+m_{\mathrm{h}})$ is
the reduced mass, and the angular momentum quantum number
$m=0,\pm1,\pm2,\ldots$ is a good quantum number.
Besides $m$, the parity $\pi = \pi_{z_\mathrm{e}} \pi_{z_\mathrm{h}}$,
related to the simultaneous exchange of $z_{\mathrm{e}}\to-z_{\mathrm{e}}$
and $z_{\mathrm{h}}\to -z_{\mathrm{h}}$, is also an exact quantum number.
This parity is even if the sum of the subband quantum numbers $i+j$ is
even, and odd otherwise.
The screened Coulomb potential with the dielectric permittivity
$\epsilon=7.5$ couples the electron-hole motion, giving rise to the
exciton.
The Coulomb potential is invariant under the parity $\pi$, but breaks
the individual parities $\pi_{z_\mathrm{e}}$ and $\pi_{z_\mathrm{h}}$.

Equation~\eqref{eq:Hamiltonian}, together with appropriate boundary
conditions, forms the eigenvalue problem $H\psi=E\psi$, which for
arbitrary $L$ can only be solved numerically due to the
three-dimensional Coulomb coupling which, in contrast to
Refs.~\cite{Ziemkiewicz2020,Ziemkiewicz2021a,Ziemkiewicz2021b,Ziemkiewicz2021c},
does not allow separation of variables.
To solve the problem, we employ an effective numerical method which is
based on the expansion of the wavefunction over a basis of $p${\,}th-order
B-splines~\cite{DeBoor,Bachau2001}
\begin{equation}
  \psi(z_{\mathrm{e}},z_{\mathrm{h}},\rho)
  = \frac{1}{\sqrt{\rho}} \sum_{ijk} c_{ijk}
  B^{p}_{i}(z_{\mathrm{e}}) B^{p}_{j}(z_{\mathrm{h}}) B^{p}_{k}(\rho) \, .
\label{eq:BsplineExpansion}
\end{equation}
B-splines are piecewise polynomial functions that have a compact
support.
This simplifies the implementation of the boundary
conditions. Furthermore, the expansion of the wavefunction over
B-splines results in a sparse and banded matrix.
The generalized eigenvalue problem for the coefficients $c_{ijk}$
is solved using ARPACK routines~\cite{arpackuserguide} which
extensively exploit the band structure of the Hamiltonian matrix.
The routines also allow us to calculate eigenvectors and thus to
extract the exciton wavefunction in coordinate space.

Depending on the boundary conditions, the expansion~\eqref{eq:BsplineExpansion}
makes it possible to calculate both bound states as well as resonances.
The bound electron-hole states of even parity can only have energies
below the lowest scattering threshold $E_{1,1}$ defined by the sum of
the lowest quantization energies $E_{\mathrm{e}1} + E_{\mathrm{h}1}$.
Above this threshold there are bound states of odd parity up to the
lowest odd-parity threshold as well as resonant states of even and odd
parities.
The resonances, that is, the quasibound states in the continuum, are
characterized by additional nonradiative broadenings due to their
coupling to the continuum of lower subbands~\cite{Landau}.
In an experiment, this additional broadening significantly smears the
sharp spectral Lorentz peaks from only the radiative recombination~\cite{ivchenko}.
These resonances can be theoretically studied by the
complex-coordinate-rotation and stabilization
methods~\cite{Scheuler2024}.
Although the complex-coordinate rotation leads to a non-Hermitian
eigenvalue problem, the stabilization method allows one to estimate
the nonradiative decay rates from solutions of the Schrödinger
equation for real observables.
By setting the wavefunction to zero at some large enough value $\rho_{\max}$,
far away from the interaction domain, we can obtain the wavefunctions
of the electron-hole resonances and, in particular, BICs as well as
their partner states.

The distinguishing feature of BICs is the absence of nonradiative
linewidth broadening and hence their giant nonradiative lifetime~\cite{Hsu2016}.
This leads to much sharper spectral peaks of BICs, broadened only by
the radiative electron-hole recombination as well as, in an
experiment, by interaction with phonons~\cite{Schweiner16a} which is
beyond the scope of this article.
However, in contrast to an ordinary resonance or, in particular, the
partner state, the wavefunction of the BIC indeed vanishes
exponentially as $\rho\to\infty$.
The derivative of the wavefunction at the origin of the coordinate
system defines the oscillator strength or the radiative decay rate of
a certain electron-hole resonance, as we will show in Sec.~\ref{subsec:oscillatorTheory}.
As was reported earlier~\cite{Aslanidis2025}, the BICs of
electron-hole pairs and their partner states appear in such a system
as a result of the destructive interference of the electron-hole
resonances associated to different subbands of the same
symmetry~\cite{Fri85b}.
Therefore, the BICs appear at much higher energies than the lowest
threshold $E_{1,1}$, namely in an energy region close to the upper
thresholds of the same symmetry.
In Ref.~\cite{Aslanidis2025} we have shown that for cuprous oxide the
BICs and their partner states appear slightly below the threshold
$E_{2,2} = E_{\mathrm{e2}} + E_{\mathrm{h2}}$.

The exciton wavefunctions depend on three coordinates, viz.\
$\rho,z_\mathrm{e},z_\mathrm{h}$ or, alternatively, $\rho$ and the
center-of-mass and relative coordinates
\begin{align}
  Z=\frac{m_{\mathrm{e}} z_{\mathrm{e}}+
  m_{\mathrm{h}}z_{\mathrm{h}}}{m_{\mathrm{e}}+m_{\mathrm{h}}}\, ,\quad
  z_\mathrm{rel}=z_{\mathrm{e}}-z_{\mathrm{h}}.
\label{eq:COMrel}
\end{align}
In the latter coordinates the Hamiltonian~\eqref{eq:Hamiltonian} reads
\begin{align}
  H &=  E_\mathrm{g} - \frac{\hbar^2}{2M}\frac{\partial^2}{\partial Z^{2}} - \frac{\hbar^2}{2\mu}\frac{\partial^2}{\partial z_\mathrm{rel}^2}-\frac{\hbar^2}{2\mu} \left( \frac{\partial^2}{\partial \rho^2}+ \frac{1}{\rho} \frac{\partial}{\partial \rho} - \frac{m^2}{\rho^2}\right)\nonumber\\ 
    & - \frac{e^2}{4\pi \epsilon_0 \epsilon \sqrt{\rho^2+z_\mathrm{rel}^2}} + V_\mathrm{e}\left(Z+\frac{m_\mathrm{h}}{m_\mathrm{e}+m_\mathrm{h}}z_\mathrm{rel}\right) \nonumber\\
    & +V_\mathrm{h}\left(Z-\frac{m_\mathrm{e}}{m_\mathrm{e}+m_\mathrm{h}}z_\mathrm{rel}\right).
\label{eq:HamiltonianZzrel}
\end{align}
Here, we explicitly see the quantization of the exciton
$(M=m_{\mathrm{e}}+m_{\mathrm{h}})$ center-of-mass motion along the $Z$
direction in the QW.
Such an effect dominates in wide QWs, i.e., for potentials which
are far away from the interaction domain and thus exhibiting weak
coupling between $Z$ and $z_{\mathrm{rel}}$~\cite{Belov2019}.
This representation of the Hamiltonian is reminiscent of
the concept of the ``dead-layer''~\cite{DAndrea1982,Tredicucci1993}.
The QW barriers $V_{\mathrm{e}}$ and $V_{\mathrm{h}}$ form a
parallelogram in the plane $(z_{\text{rel}},Z)$.
This concept was used to phenomenologically describe the effect of the
QW barriers, i.e., a suppression of the exciton wavefunction in the
corners of this parallelogram.
The wavefunctions and symmetries of the exciton center-of-mass
quantized states are defined by expressions similar to
Eq.~\eqref{eq:1Dconfinmentfunctions}.

In order to properly visualize the three-dimensional wavefunction in
two dimensions one has to choose an appropriate representation.
To this end, we either slice the wavefunction $\psi( \rho,z_\mathrm{rel}, Z)$ 
at a fixed value of $\rho=\rho_{0}$ or $Z=Z_{0}$, e.g.,
\begin{align}
  \psi(\rho,z_{\mathrm{rel}}) = \psi( \rho, Z=Z_{0},z_{\mathrm{rel}}) \, ,
\label{eq:rhozrel}
\end{align}
or we integrate it over the radial variable to show a behavior of the
wavefunction along the confinement directions,
\begin{align}
  \psi(z_\mathrm{e},z_\mathrm{h})
  = \int_{0}^{\rho_{\max}} |\psi(\rho,z_\mathrm{e},z_\mathrm{h})|^2 \rho \, \mathrm{d}\rho.
\label{eq:psizezh}
\end{align}
It is worth noting that the latter representation is similar to the
definition of a probability density, and thus is always positive.
As a result, it can obscure the parity of the wavefunction.

%%%%%%%%%%%%%%%%%%%%%%%%%%%%%%%%%%%%%%%%%%%%%%%%%%%%%%%%%%%%%%%%%%%%%%%%%%%%%%%%%
\subsection{Oscillator strengths}
\label{subsec:oscillatorTheory}
The oscillator strength $f$ measures the probability of absorption or
emission of electromagnetic radiation such as light by an electron as
it transitions between energy levels in, e.g., an atom or a molecule.
It is directly related to the dipole moment matrix element of the transition.
In atomic physics it can be calculated by Fermi's Golden Rule~\cite{Landau}
\begin{align}
  f_{\mathrm{i}\rightarrow \mathrm{f}}
  = \frac{2m_{\mathrm{e}}}{\hbar^2} \frac{E_{\mathrm{f}}-E_{\mathrm{i}}}{e^2}
  \left|\bra{\psi_{\mathrm{f}}} \boldsymbol{r} \ket{\psi_{\mathrm{i}}} \right|^2
\label{eq:atomic}
\end{align}
and represents transitions from an initial state $\mathrm{i}$ to the
final state $\mathrm{f}$.
While excitons behave in many ways like a hydrogenlike two-particle
system, it is important to keep in mind that they are actually excited
states of the crystal.
Selection rules for optical transitions and oscillator strengths are
influenced by the symmetry properties and parities of the valence and
conduction bands involved.
We thus need to render our description of excitons more precise to
take this into account.
A more general examination of this problem was undertaken in
Ref.~\cite{Schweiner17c} for bulk states.
We will follow similar ideas in this section to derive the radiative
characteristics of excitons in cuprous oxide QWs, but, contrary to
Ref.~\cite{Schweiner17c}, use a hydrogenlike envelope function.

Let $| \Phi_0 \rangle$ be the ground state of the semiconductor, where
all electrons are in the fully occupied valence bands.
An exciton state in this many-particle framework is given by
\begin{equation}
  |\psi^{\sigma \tau}_{vc,\nu} \rangle = \sum_{\boldsymbol{q}_c,\boldsymbol{q}_v} f^{\tau,\sigma}_{vc,\nu}(\boldsymbol{q}_c,\boldsymbol{q}_v) c^\dagger_{c,\boldsymbol{q}_c,\tau} c_{v,\boldsymbol{q}_v,\sigma} | \Phi_0 \rangle
\label{eqref:ExcitonManyBody}
\end{equation}
where $c^\dagger_{c,\boldsymbol{q}_c}$ creates an electron
with crystal momentum $\boldsymbol{q}_c$ and electron spin
$\tau$ in the conduction band $c$, and $c_{v,\boldsymbol{q}_v}$
creates a hole with crystal momentum $-\boldsymbol{q}_v$ and
hole spin $-\sigma$ in the valence band $v$.
The envelope function 
$f^{\tau,\sigma}_{vc,\nu}(\boldsymbol{q}_{c},\boldsymbol{q}_{v})$,
is the solution of the hydrogenlike Hamiltonian~\eqref{eq:Hamiltonian},
Fourier transformed into crystal-momentum space.
The index $\nu$ represents the quantum numbers of the exciton, and
$\tau, \sigma, c, v$ are taken as good quantum numbers for simplicity.
More general states are obtained by appropriate superpositions.

In the physics of excitons we are mainly interested not in transitions
between particular exciton states~\cite{PhysRevB.104.085204}, but in
the excitations of the crystal from its ground state 
$\Phi_0$ into an exciton state $\psi^{\sigma\tau}_{vc,\nu}$~\cite{Rashba}.
Therefore, Eq.~\eqref{eq:atomic} has to be properly revised.
It is worth noting that the energy $E_\mathrm{g} = 2.17208$~eV in the
Hamiltonian~\eqref{eq:Hamiltonian} is the gap energy between the
maximum of the valence band and the minimum of the conduction band.
This energy and the dielectric constant $\epsilon = 7.5$ determine
the approximate wavelength $\lambda \approx 200$~nm in the crystal for
the excitonic transition of the yellow series.
This wavelength is large compared to the size of excitons considered
in our paper and thus justifies using the long wavelength approximation.
Assuming separable states, so that we have a notion of true exciton,
photon, and phonon states, we can treat the exciton-photon interaction
as a perturbation.
We do not consider polariton effects described, e.g., in Ref.~\cite{Stolz2018}.

In the bulk, the total exciton momentum $\boldsymbol{K}$ is conserved.
The transition probability from the crystal ground state $\Phi_0$ to
an exciton state $\psi^{\sigma \tau}_{vc,\nu \boldsymbol{K}}$ is
proportional to the squared matrix element $|M|^2$, where
\begin{equation}
  M = \left\langle \psi^{\sigma \tau}_{vc, \nu \boldsymbol{K}} \left| -\frac{e}{m_0} \boldsymbol{A}_0(\boldsymbol{\kappa}, \xi) \sum_{j=1}^{N} e^{i\kappa \cdot \boldsymbol{r}_j} \boldsymbol{p}_j \right| \Phi_0 \right\rangle.
\end{equation}
Here, $\boldsymbol{A}_0(\kappa, \xi) = {A}_0(\kappa,\xi)\boldsymbol{e}_\xi^\kappa$
is the amplitude of the vector potential of the radiation field,
$\boldsymbol{\kappa}$ is the wave vector, and $\xi$ is the polarization.
There are $N$ electrons in total in the valence and conduction bands
and the operator $\boldsymbol{p}_j$ denotes the momentum of the $j$-th electron.
The state $\psi^{\sigma \tau}_{vc, \nu \boldsymbol{K}}$ is constructed
from the Bloch functions and the envelope function analogously to
Eq.~\eqref{eqref:ExcitonManyBody}.
The matrix element $M$ can be expressed in terms of the Bloch
functions and the Fourier coefficients $f^{\sigma
  \tau}_{vc,\nu}(\boldsymbol{q})$ of the exciton wavefunction as in
Eq.~\eqref{eqref:ExcitonManyBody}, where we can now restrict ourselves
to the relative momentum $\boldsymbol{q}$ between the electron and hole.
Using the $\boldsymbol{k} \cdot \boldsymbol{p}$ perturbation theory, the
expression for the matrix element is expanded to first order in
$\boldsymbol{K}$ and $\boldsymbol{q}$.
This gives
\begin{align}
  M  &= -\frac{e}{m_0^2} \boldsymbol{A}_0 (\kappa, \xi) N \delta_{\tau \sigma} \delta_{\boldsymbol{\kappa} \boldsymbol{K}}  \sum_{\boldsymbol{q}} f^*_{vc,\nu}(\boldsymbol{q}) \bigg[ 
       \left\langle u_{c0} \left| \boldsymbol{e}_\xi^\kappa \cdot \boldsymbol{p} \right| u_{v0} \right\rangle  \nonumber\\ 
     & + \left\langle u_{c0} \left| (\boldsymbol{e}_\xi^\kappa \cdot \boldsymbol{p}) M_v\big(\boldsymbol{p} \cdot (\boldsymbol{q}-\frac{m_\mathrm{h}}{m_\mathrm{e}+m_\mathrm{h}}\boldsymbol{K})\big) 
       \right| u_{v0} \right\rangle \nonumber\\ 
     & + \left\langle u_{c0} \left| \big((\boldsymbol{q}+\frac{m_\mathrm{e}}{m_\mathrm{e}+m_\mathrm{h}}\boldsymbol{K})\cdot\boldsymbol{p} \big) M_c\big( \boldsymbol{p}\cdot \boldsymbol{e}_\xi^\kappa\big) 
       \right| u_{v0} \right\rangle \bigg],
\label{eq:M}
\end{align}
where $u_{c0}$ and $u_{v0}$ are Bloch functions in the respective
bands at the $\Gamma$ point, and ${M}_v$ and ${M}_c$ are projection
operators in the space of Bloch functions~\cite{Schweiner17c}.
The operator $\boldsymbol{p}$ here denotes the momentum operator as it is
applied to single-electron states.
In Cu$_2$O, the parity of the valence and conduction bands results in
the vanishing of the first term.
Therefore, only higher order terms in $\boldsymbol{q}$ and $\boldsymbol{K}$
contribute to the oscillator strength.
From these, taking into account the symmetry properties of the bands,
the following expression for the relative oscillator strength
$f^{\text{rel}}_{\xi \nu \boldsymbol{K}}$ without quadrupole terms can be
obtained~\cite{Schweiner17c},
\begin{equation}
  f^{\text{rel}}_{\xi \nu \boldsymbol{K}} = \left| \lim_{r \to 0} \frac{\partial}{\partial r} \left\langle T^D_{\xi \boldsymbol{K}} | \psi_{\nu \boldsymbol{K}} \right\rangle \right|^2,
\end{equation}
where the overlap with $\big\langle T^D_{\xi \boldsymbol{K}}\big|$ determines the
strength of absorption for different polarizations.
The precise form of these $\big\langle T^D_{\xi \boldsymbol{K}}\big|$ in the case
of a bulk crystal can be found in Ref.~\cite{Schweiner17c} and results from
group-theoretical considerations in the context of the cubic $O_{\mathrm{h}}$ group.
Only relative oscillator strengths can be calculated, as the absolute
values of certain prefactors cannot be determined.
For bulk states in cuprous oxide the oscillator strength for the P-excitons 
is proportional to
\begin{align}
  f_{\mathrm{rel}}(n) \approx \frac{n^2-1}{3\pi a_{B,ex}^5 n^5} \propto \frac{1}{n^3},
\label{eq:scaling}
\end{align}
which is proportional to the derivative of the exciton envelope
wavefunction at vanishing separation between electron and
hole~\cite{Assmann20}.

At this point, we can discuss the calculations for QW exciton states.
The main difference to the bulk calculations, given in detail in
Ref.~\cite{Schweiner17c}, is that the wave vector $\boldsymbol{K}$ is
no longer a good quantum number due to the confinement along the $z$ axis.
However, the two-dimensional wave vector $\boldsymbol{K}_\mathrm{2D}$
in the $xy$ plane is still conserved.
We start the derivation from the formula for the matrix element
$M$~\cite{Schweiner17c}
\begin{align}
  M &= -\frac{e\hbar}{m_0^2} A_0(\boldsymbol{\kappa},\xi)
  \sqrt{N} \delta_{\tau,\sigma}\delta_{\boldsymbol{\kappa},\boldsymbol{K}}\nonumber \\
  &\times\lim \limits_{\boldsymbol{r}\to 0}(\tilde{\boldsymbol{M}}_v
    +\tilde{\boldsymbol{M}}_c)i\nabla F_{vc,\nu}^*(\boldsymbol{r}) \, .
\label{eq:M1}
\end{align}
To simplify this equation we rewrite
\begin{align}
  (\tilde{\boldsymbol{M}}_v +\tilde{\boldsymbol{M}}_c)i\nabla F_{vc,\nu}^*(\boldsymbol{r}) = i(\hat{M}_v+\hat{M}_c)
\end{align}
with
\begin{subequations}
\begin{align}
  \hat{M}_v  &= \bra{u_{c\boldsymbol{0}}} (\hat{\boldsymbol{e}}_{\xi\boldsymbol{K}}\cdot \boldsymbol{p})M_v(\nabla F_{vc,\nu}^*(\boldsymbol{r}) \cdot \boldsymbol{p} )\ket{u_{v\boldsymbol{0}}}, \\
  \hat{M}_c  &= \bra{u_{c\boldsymbol{0}}} (\nabla F_{vc,\nu}^*(\boldsymbol{r}) \cdot \boldsymbol{p}) M_c (\hat{\boldsymbol{e}}_{\xi\boldsymbol{K}} \cdot \boldsymbol{p}) \ket{u_{v\boldsymbol{0}}} .
\end{align}
\end{subequations}
Cuprous oxide has the crystal symmetry $O_\mathrm{h}$.
The ground state Bloch function of the valence band
$u_{v\boldsymbol{0}}$  transforms according to the irreducible
representation $\Gamma^+_5$ and the conduction band
$u_{c\boldsymbol{0}}$ transforms according to $\Gamma^+_1$.
The operators $M_c$ and $M_v$ are projection operators.
These operators have to transform according to the irreducible
representation $\Gamma^+_1$ of $O_\mathrm{h}$, due to their symmetry.
The momentum operator $\boldsymbol{p}$ transforms according to $\Gamma^-_4$.
The matrix element is only non-zero if the product representation of
the terms between bra and ket transforms as the representation of
$u_{v\boldsymbol{0}}$.
Alternatively, we can regard a fixed polarization
$\hat{\boldsymbol{e}}_{\xi\boldsymbol{K}}$ and require that the
product representation of the gradient
$(\nabla F_{vc,\nu}^*(\boldsymbol{r}) \cdot \boldsymbol{p})$ and the
Bloch function $u_{v\boldsymbol{0}}$ transforms as 
$(\hat{\boldsymbol{e}}_{\xi\boldsymbol{K}}\cdot\boldsymbol{p})$.
The coupling coefficients for the product representation 
$\Gamma_4 \times \Gamma_5 \rightarrow \Gamma_4$ are given in
Ref.~\cite{koster_properties_1963} as
\begin{subequations}
\begin{align}
	\frac{1}{\sqrt{2}}(a_y^4 a_{xy}^5+a_z^4a_{zx}^5) &\sim b_x^4\, , \\
	\frac{1}{\sqrt{2}}(a_z^4 a_{yz}^5+a_x^4a_{xy}^5) &\sim b_y^4\, , \\
	\frac{1}{\sqrt{2}}(a_x^4 a_{zx}^5+a_y^4a_{yz}^5) &\sim b_z^4\, .
\end{align}
\end{subequations}
Here $a^4_i$, $b^4_i$ for $i$ = $x$, $y$, $z$ denote basis functions
of the $\Gamma_4^-$ representation, with the $a^4_i$ corresponding to
the components of the gradient.
Analogously, $a^5_i$ with $i$ = $yz$, $xz$, $xy$ are basis functions
of the representation $\Gamma^+_5$ corresponding to the valence band
Bloch functions.
Following this scheme, the resulting product representation components
$b^4_i$ must correspond to the analogous basis functions of the
polarization.
Using this condition, we can see that the transition matrix elements
have to be
\begin{subequations}
\begin{align}
  M &\propto (\partial_y\bra{u_{xy\boldsymbol{0}}} +
      \partial_z\bra{u_{zx\boldsymbol{0}}})\ket{\psi^{\sigma
      \tau}_{vc,\nu \boldsymbol{K}_\mathrm{2D}}},\\
  M &\propto (\partial_z\bra{u_{yz\boldsymbol{0}}} +
      \partial_x\bra{u_{xy\boldsymbol{0}}})\ket{\psi^{\sigma
      \tau}_{vc,\nu \boldsymbol{K}_\mathrm{2D}}},\\
  M &\propto (\partial_x\bra{u_{zx\boldsymbol{0}}} +
      \partial_y\bra{u_{yz\boldsymbol{0}}})\ket{\psi^{\sigma
      \tau}_{vc,\nu \boldsymbol{K}_\mathrm{2D}}},
\end{align}
\label{eq:TransitionRulesPolarization}
\end{subequations}
for polarizations along the $x$, $y$, and $z$ axis, respectively.

In the QW, the center-of-mass momentum along the $z$ axis
$K_z$ is not a good quantum number.
However, the wavefunction can be written as a superposition of states
with well-defined $K_z$ as
\begin{widetext}
\begin{align}
	\ket{\psi^{\sigma \tau}_{vc,\nu \boldsymbol{K}_\mathrm{2D}}} &= \left[ \iiint \psi^{\sigma \tau}_\nu(\rho,z_{\mathrm{rel}},Z) \ket{\rho,z_\mathrm{rel},Z} \rho \mathrm{d}\rho \ \mathrm{d}z_\mathrm{rel} \mathrm{d}Z\right]\ket{m}\ket{\boldsymbol{K}_{2D}}\nonumber \\
	& = \left[ \iiint\int\psi^{\sigma \tau}_\nu(\rho,z_{\mathrm{rel}},Z) \frac{e^{im\phi}}{\sqrt{2\pi}}\ket{\rho,\phi,z_\mathrm{rel},Z} \rho \ \mathrm{d}\rho \mathrm{d}\phi \mathrm{d}z_\mathrm{rel} \mathrm{d}Z\right]\ket{\boldsymbol{K}_{2D}} \nonumber\\
	&= \iiint \int \psi^{\sigma \tau}_\nu(\rho,z_{\mathrm{rel}},Z) \frac{e^{im\phi}}{\sqrt{2\pi}}  \left(\frac{1}{\sqrt{L}}\int e^{-i K_z Z}\ket{\rho,\phi,z_\mathrm{rel},K_z} \mathrm{d}K_z\right) \rho \ \mathrm{d}\rho \ \mathrm{d}\phi \ \mathrm{d}z_\mathrm{rel} \ \mathrm{d}Z\ket{\boldsymbol{K}_{2D}}\nonumber \\
	&= \iiint\int \left( \int \psi^{\sigma \tau}_\nu(\rho,z_\mathrm{rel},Z) \frac{e^{im\phi}}{\sqrt{2\pi L}}e^{-iK_z Z} \mathrm{d}Z \right) \ket{\boldsymbol{K},\rho,\phi,z_\mathrm{rel}} \rho \ \mathrm{d}\rho \ \mathrm{d}\phi \ \mathrm{d}z_\mathrm{rel} \ \mathrm{d}K_z.
\label{eq:StateSuperpositionKz}
\end{align}
\end{widetext}
The expression in round brackets in the last line of
Eq.~\eqref{eq:StateSuperpositionKz} corresponds to the
Fourier-transformed envelope function $F_{vc,\nu}$ \cite{Schweiner17c}
for a component with well-defined total exciton momentum
$\boldsymbol{K}$, which is a good quantum number in the bulk.
In the QW it can be split as
$\boldsymbol{K} = \boldsymbol{K}_{2D} + K_z\boldsymbol{e}_z$, 
where only $\boldsymbol{K}_{2D}$ is a good quantum number, but not $K_z$.
The term $\delta_{\boldsymbol{\kappa},\boldsymbol{K}}$ in the transition
matrix element~\eqref{eq:M1}, in the strong dipole approximation
$\kappa \approx 0$, picks out the term with vanishing $K_z = 0$. 
The projection operators and derivatives in
Eqs.~\eqref{eq:TransitionRulesPolarization} then need to be applied to
this term.

As in the bulk, we will also neglect common prefactors as we are
interested in relative oscillator strengths, rather than their
absolute values.
When considering a polarization in the $xy$ plane, we have to calculate
$(\partial_y\bra{u_{xy\boldsymbol{0}}} + \partial_z\bra{u_{zx\boldsymbol{0}}}) \left( \psi^{\sigma \tau}_\nu(\rho,z_\mathrm{rel},Z) e^{-im\phi}\right)$ and 
$(\partial_z\bra{u_{yz\boldsymbol{0}}} + \partial_x\bra{u_{xy\boldsymbol{0}}}) \left( \psi^{\sigma \tau}_\nu(\rho,z_\mathrm{rel},Z) e^{-im\phi}\right)$, for $x$ and $y$ polarizations, respectively.

Let us now consider the derivatives in polar coordinates,
\begin{subequations}
\begin{align}
  &\frac{\partial}{\partial x} (\psi^{\sigma \tau}_\nu e^{i m \phi})
  =  \left(\cos{\phi} \frac{\partial}{\partial \rho} - \frac{1}{\rho} \sin{\phi} \frac{\partial}{\partial \phi}\right)\psi^{\sigma \tau}_\nu e^{i m \phi}  \nonumber \\
  &=  \left(\cos{\phi} \frac{\partial\psi^{\sigma
    \tau}_\nu(\rho,z_\mathrm{rel},Z)}{\partial \rho} - i m
    \frac{\psi^{\sigma \tau}_\nu}{\rho} \sin{\phi} \right) e^{i m \phi} , \\
  &\frac{\partial}{\partial y} (\psi^{\sigma \tau}_\nu e^{i m \phi})
  =  \left(\sin{\phi} \frac{\partial}{\partial \rho} + \frac{1}{\rho} \cos{\phi} \frac{\partial}{\partial \phi}\right)\psi^{\sigma \tau}_\nu e^{i m \phi}  \nonumber \\
  &=  \left(\sin{\phi} \frac{\partial\psi^{\sigma \tau}_\nu(\rho,z_\mathrm{rel},Z)}{\partial \rho} + i m \frac{\psi^{\sigma \tau}_\nu}{\rho} \cos{\phi} \right) e^{i m \phi} .
\end{align}
\end{subequations}
We require that the derivative is continuous at the origin.
It is evident that this can only result in non-zero values if $m = \pm
1$.
In this case, $\psi^{\sigma \tau}_\nu(\rho = 0,z_\mathrm{rel},Z) = 0$, and thus
$\lim_{\rho \rightarrow 0} \psi^{\sigma \tau}_\nu(\rho,z_\mathrm{rel},Z)/ \rho 
= \partial_\rho \psi^{\sigma \tau}_\nu(\rho,z_\mathrm{rel},Z) $.
We thus obtain the result
\begin{align}
  \frac{\partial}{\partial x} \left( \psi_\nu(\rho,z_\mathrm{rel},Z) e^{\pm i\phi}\right) &= \frac{\partial \psi_\nu}{\partial \rho} \nonumber, \\
  \frac{\partial}{\partial y} \left( \psi_\nu(\rho,z_\mathrm{rel},Z) e^{\pm i\phi}\right) &= \pm i \frac{\partial \psi_\nu}{\partial \rho} \nonumber.
\end{align}
We also require continuity of the derivative along the $z$ direction
at the origin, $\partial_z(\psi_\nu(\rho,z_\mathrm{rel},Z) e^{im\phi})$.
In this case, the derivative can only be non-zero for $m=0$.
If we restrict ourselves to $m = \pm 1$, the $z$ derivatives vanish.

From this, we can deduce selection rules for $m$ based on the
derivatives appearing in
Eqs.~\eqref{eq:TransitionRulesPolarization}. We see that light
polarized along $x$ and $y$ can excite both $m=0$ and $m=1$ states,
whereas light polarized along $z$ can only excite $m=1$ states.
This is markedly different from the situation in atomic physics.
The cause of this discrepancy is that the description of electron and
hole has to include the Bloch functions in the respective bands, so
$m$ does not represent the total angular momentum of the exciton
itself.
This is taken into account by the projection operators in
Eqs.~\eqref{eq:TransitionRulesPolarization}.

For the calculation of oscillator strengths, along with the
derivatives in $x$ and $y$ direction, we then need to project into the
subspace given by $\ket{u_{xy\boldsymbol{0}}}\ket{S=0,M_S =0}$ with
the total spin
$\boldsymbol{S} = \boldsymbol{S}_\mathrm{e}+\boldsymbol{S}_\mathrm{h}$, 
which can be done using Clebsch-Gordan coefficients.
This condition takes into account that spin flips are forbidden in
electric dipole transitions.
In the hydrogenlike model for the envelope function used here, taking
these conditions on the spin degrees of freedom into account will only
lead to a constant coefficient that is the same for all states in the
yellow series which does not affect relative oscillator strengths.
Note that, however, in a model that includes the non-parabolicity of
the valence bands, these conditions would be non-trivial.
One way of treating these projections would be similarly as discussed
in Ref.~\cite{Schweiner17c}, introducing a quasispin $I$ to model the
valence band states.

By choosing circular polarization 
$\hat{\boldsymbol{e}}_{\xi\boldsymbol{K}} = \hat{\boldsymbol{e}}_x \pm i\hat{\boldsymbol{e}}_y$
we get
\begin{align}
  M \propto \lim \limits_{\rho \to 0} \int \frac{\partial}{\partial \rho} \psi(\rho,z_\mathrm{rel}
    =0, Z) \ \mathrm{d}Z \,
\end{align}
and therefore 
\begin{align}
\label{eq:osci}
  f_{\mathrm{rel}} &\propto  \left| \lim \limits_{\rho \to 0} \int \frac{\partial}{\partial \rho} 
    \psi(\rho,z_\mathrm{rel}=0, Z) \ \mathrm{d}Z \right|^2 \nonumber \\
	&= \left| \lim \limits_{\rho \to 0} \int \frac{\psi(\rho,z_\mathrm{rel}=0, Z)}{\rho} \mathrm{d}Z 
    \right|^2.
\end{align}
The second expression in Eq.~\eqref{eq:osci} is more convenient for
numerical evaluations.
Technical remarks on calculating the limit $\rho$ $\rightarrow$ $0$
are given in Appendix.
Note that the $Z$ integral in Eq.~\eqref{eq:osci} is carried out along
the laser beam direction indicated by the red arrow in Fig.~\ref{fig:QW_sketch}.

%%%%%%%%%%%%%%%%%%%%%%%%%%%%%%%%%%%%%%%%%%%%%%%%%%%%%%%%%%%%%%%%%%%%%%%%
\section{Results and Discussion}
\label{RandD}
\subsection{Exciton energies}
\label{sec:energies}
\begin{figure}
  \includegraphics[width = \columnwidth]{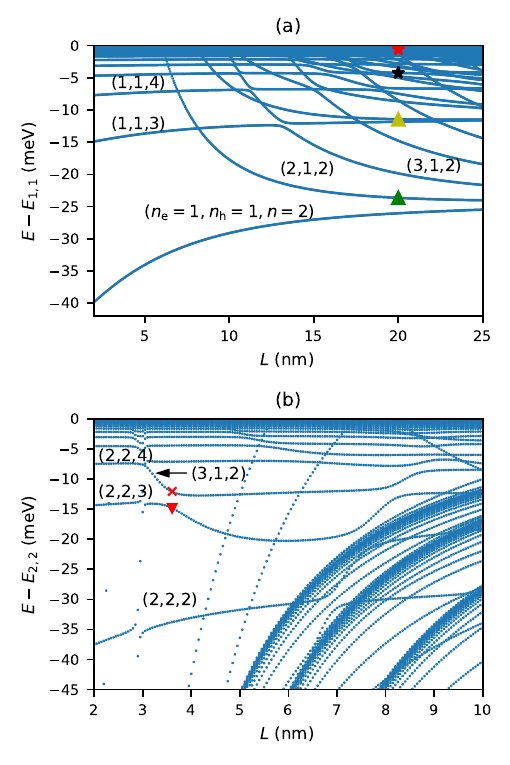}
  \caption{Energy spectra for exciton states with $m=1$ over the QW
    width $L$ with respect to the energy thresholds $E_{1,1}$ (a) and
    $E_{2,2}$ (b), respectively. The marked states are discussed in
    the text. Some states are marked by approximate quantum numbers
    $(n_\mathrm{e},n_\mathrm{h},n)$.}
\label{fig:spectra}
\end{figure}

As it was shown recently~\cite{Belov2024}, excitons in cuprous oxide
QWs have a very rich energy spectrum.
Two examples of such spectra for angular momentum $m=1$ are depicted in 
Fig.~\ref{fig:spectra}.
They show the dependence of energy levels of the electron-hole states
on the QW width.
In Fig.~\ref{fig:spectra}(a) the energy curves are shown with
respect to the lowest scattering threshold given by the sum of the
quantum-confinement energies $E_{1,1} = E_{\mathrm{e1}} + E_{\mathrm{h1}}$.
Thus, these states correspond to bound electron-hole states, i.e.\ excitons. 
One observes that for narrow QWs there is only one 2D Rydberg series,
however, more energy levels from the upper quantum confinement
subbands appear as the QW thickness increases.
Some of the energy curves are labeled by approximate quantum
numbers $(n_{\mathrm{e}},n_{\mathrm{h}},n)$ of electron and hole
confinement subbands as well as the principal quantum number of the
hydrogenlike system.
Furthermore, at QW width $L=20\,$nm some states whose wavefunctions we
will discuss in Sec.~\ref{sec:wavefunctions} are marked by colored symbols.

The density of bound states increases when approaching the scattering
threshold, in particular for wide QWs where many Rydberg states as well
as lower states from upper subbands show crossings and avoided crossings.
These features of the spectrum indicate the crossover between the
strong and the weak confinement regimes.
As different states have different spatial extent, the crossover depends 
on the particular state.
In the limiting cases of extremely narrow and wide QWs, the energy levels 
are perfectly aligned according to their exact quantum numbers.
In the crossover region, the energy levels are being rearranged and
thus interact due to the Coulomb coupling and exhibit crossings and
avoided crossings.

\begin{figure}
  \includegraphics[width = \columnwidth]{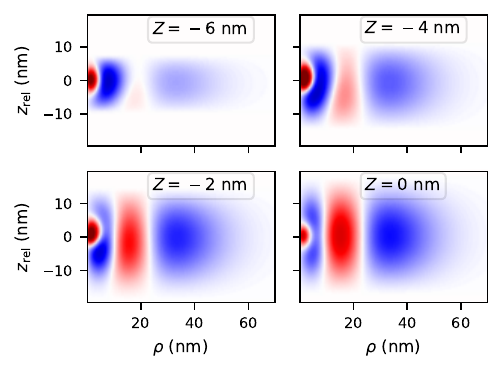}
  \caption{Exciton wavefunction $\psi(\rho,z_\mathrm{rel},Z)$ with
    $m=1$ and approximate quantum numbers $(1,1,5)$ in a QW with width
    $L=20\,$nm. In each figure a different center-of-mass coordinate $Z$ is
    shown. Changing the sign of $Z$ reflects the wavefunction along
    the $z_{\mathrm{rel}} = 0$ line.}
\label{fig:n5}
\end{figure}
The BICs and their partner states, as particular types of
electron-hole resonances, appear at much higher energies, closer to
the threshold $E_{2,2}$ than to the lowest threshold at $E_{1,1}$~\cite{Aslanidis2025}.
In Fig.~\ref{fig:spectra}(b) the energy curves of electron-hole
resonances are depicted with respect to the scattering threshold $E_{2,2}$.
Here, we explicitly show one such pair of states which emerges
about 200~meV above the lowest scattering threshold $E_{1,1}$.
In particular, the energy curves of interacting resonances with quantum
numbers $(2,2,3)$ and $(3,1,2)$ exhibit the avoided crossing for QW widths $L=3$--$4\,$nm.
At $L=3.606\,$nm these resonances produce the BIC (indicated by the
red triangle) and its partner state (indicated by the red cross) with
nonzero broadening.
The nearly vertical dotted curves at $L<7\,$nm are artificial
discretized continuum states due to rigid wall boundary conditions at
large $\rho=\rho_{\mathrm{max}}$.

%%%%%%%%%%%%%%%%%%%%%%%%%%%%%%%%%%%%%%%%%%%%%%%%%%%%%%%%%%%%%%%%%%%%%%%%%%%%%%%
\subsection{Wavefunctions}
\label{sec:wavefunctions}
According to Eq.~\eqref{eq:osci}, the radiative characteristics of an
exciton in a cuprous oxide QW are defined by the derivative
of the wavefunction at the origin.
In the region where many energy levels are close to each other, the
wavefunction of any particular exciton state is distorted due to the
interaction with other states.
Hence, these perturbations can affect their symmetry and radiative properties.
In order to study these effects, we plot and analyze the wavefunctions
of the excitons in QWs.
It is worth noting that the studied wavefunctions are three-dimensional.
To be able to plot them in the figures we use the representations in
Eqs.~\eqref{eq:rhozrel} and \eqref{eq:psizezh}, thereby reducing the
wavefunctions to their respective two-dimensional distributions.

\begin{figure}
  \includegraphics[width = \columnwidth]{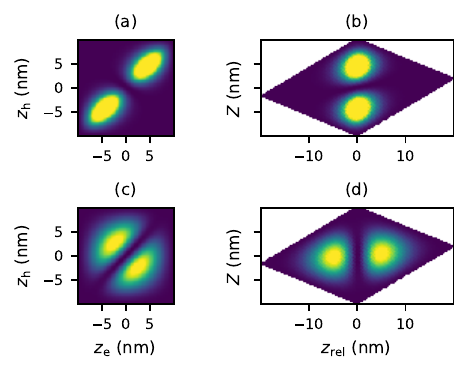}
  \caption{The wavefunction of the states with quantum numbers $(2,1,2)$ (a), (b) and
    $(1,2,2)$ (c), (d) are shown. The representation in (a) and
    (c) is according to Eq.~(\ref{eq:psizezh}), while in (b) and (d) a
    coordination transformation into the ($z_\mathrm{rel},Z$) basis
    was performed.  The QW width is $L=20\,$nm and the magnetic quantum number is $m=1$.}
\label{fig:zezh}
\end{figure}
In Fig.~\ref{fig:n5} we show the wavefunction of the Rydberg state
marked in Fig.~\ref{fig:spectra} by the black asterisk at $L=20\,$nm.
This is a state originating from the lowest electron and hole 
confinement subbands with the principal quantum number $n=5$.
The figure shows slices of the wavefunction obtained for different
center-of-mass coordinates using Eq.~\eqref{eq:rhozrel}.
The red and blue colors represent positive and negative values of the
wavefunction, respectively.
Taking into account that $m=1$, the number of knots of the
wavefunctions along the $\rho$ axis confirms the value of the
principal quantum number.
One also observes that the spatial extension along the $z_\mathrm{rel}$
axis becomes smaller for center-of-mass coordinates $Z$ closer to the
QW barriers.
Obviously, in that region the wavefunction becomes more confined.

The asymmetry of the wavefunction for large $|Z|$ observed in
Fig.~\ref{fig:n5} originates from the asymmetric positions of the
quantum confinement barriers in the $(z_{\mathrm{rel}},Z)$ plane.
To show this effect, we look at the probability distribution over
$(z_\mathrm{e},z_\mathrm{h})$ plane as defined by the
representation~\eqref{eq:psizezh} as well as at the same distribution
in the $(z_{\mathrm{rel}},Z)$ plane.
These distributions for the states marked by the green and yellow 
triangles in Fig.~\ref{fig:spectra} are shown in Fig.~\ref{fig:zezh}.
Plots of the two wavefunctions with approximate quantum numbers
$(n_{\mathrm{e}},n_{\mathrm{h}},n)=(2,1,2)$ and $(1,2,2)$ in the
$(z_\mathrm{e},z_\mathrm{h})$ plane and the $(z_{\mathrm{rel}},Z)$
plane are given in Figs.~\ref{fig:zezh}(a), (c) and
Figs.~\ref{fig:zezh}(b), (d), respectively.
Interestingly, a change of the coordinates leads to a change of the
calculation domain.
One can see in the left panels of Fig.~\ref{fig:zezh} that the QW
barriers in $(z_{e},z_{h})$ plane form an exact square, however, in
$(z_{\mathrm{rel}},Z)$ plane it is a parallelogram, see the right panels.
Its sides are inclined to the quantization axes by different angles
due to unequal values of the effective masses $m_{\mathrm{e}}$ and
$m_{\mathrm{h}}$ in Eq.~\eqref{eq:COMrel}.
The small values of $|Z|$ correspond to the diagonal of the parallelogram,
for which the asymmetry of the wavefunction is hardly visible.
However, larger $|Z|$ represent corners of the parallelogram, for which
the asymmetry is more pronounced.

The state shown in Figs.~\ref{fig:zezh}(a), (b) belongs to the first
excited quantum confinement subband of the electron and the ground
hole subband, i.e., the wavefunction has odd parity.
It is worth noting that the representation~\eqref{eq:psizezh} does not
distinguish between positive and negative values of the wavefunctions.
Nevertheless, one observes that the wavefunction is close to zero
along the antidiagonal $z_\mathrm{e}= -z_\mathrm{h}$, with two
distinct peaks along the diagonal $z_\mathrm{e} = z_\mathrm{h} $.
This is the opposite to the wavefunction of the state with quantum 
numbers $(1,2,2)$ shown in Figs.~\ref{fig:zezh}(c), (d).
This, in turn, represents the lowest state of the first excited hole
quantum confinement subband and the ground electron subband.
The different symmetries of the different quantum confinement subbands
lead to the different symmetry axes of the exciton wavefunctions.
A slight asymmetry of the shown wavefunctions in the
$(z_{\mathrm{rel}},Z)$ plane in subfigures (b) and (d) due to the
asymmetry of the QW barriers takes place as already discussed.

\begin{figure}
  \includegraphics[width = \columnwidth]{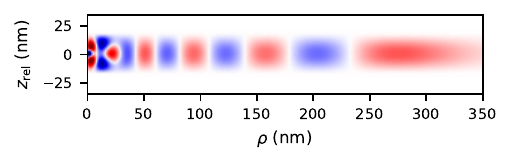}
  \caption{The visualization of the wavefunction of a highly
    excited state. The QW has the width $L = 20$\,nm and
    magnetic quantum number $m = 1$. The number of knots along
    the $\rho$ axis allows to estimate the approximate principal
    quantum number to be $n=11$. The extent along the $\rho$
    axis is around 350\,nm.}
\label{fig:small1}
\end{figure}
The wavefunction of a state with large approximate principal quantum
number marked by the red asterisk in Fig.~\ref{fig:spectra} is shown
in Fig.~\ref{fig:small1}.
From the number of knots along $\rho$ axis we can estimate the
principal quantum number $n=11$.
One also observes the excitation along the $z_{\mathrm{rel}}$ axis
which results from a mixing with neighboring states from higher
quantum confinement subbands.
This can happen because of avoided crossings.
Due to the Coulomb interaction, the wavefunction is non-separable as
can be seen from the non-trivial node structures.
However, this is only the case in the region up to $\rho=40\,$nm.
For larger values of $\rho$ one detects no influence of this effect
and only the quenching behavior of the wavefunction with an extension
up to $350\,$nm is observed.

\begin{figure}
\includegraphics[width = \columnwidth]{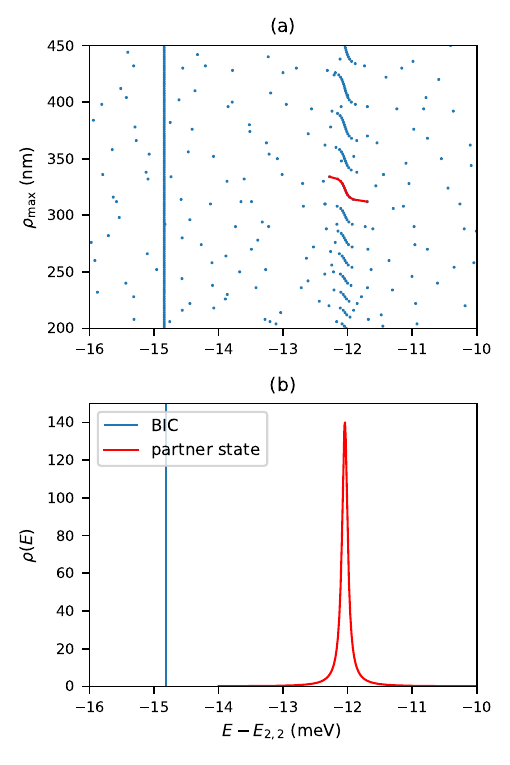}
\caption{(a) Stabilization diagram for states in a QW with
  width $L = 3.606\,$nm. The straight line indicates the
  BIC and the segments around $-12\,$meV correspond to the partner
  state.  (b) Estimated density of states for the BIC (delta function)
  and its partner state (Lorentzian shape).}
\label{fig:stabilization}
\end{figure}
In Fig.~\ref{fig:stabilization} the stabilization diagram for the BIC
and its partner state is shown.
This pair of states emerges high in the continuum, approximately
$200\,$meV above the lowest scattering threshold $E_{1,1}$.
Figure~\ref{fig:spectra}(b) shows their energies with respect to 
the threshold $E_{2,2}$ of the same symmetry as $E_{1,1}$.
Figure~\ref{fig:stabilization}(a) shows that the energy of the BIC is
independent of the position $\rho_{\max}$ of the boundary condition.
For the partner state, the energy varies slightly.
This takes place in the energy regions between the avoided crossings
of a particular resonance with the discretized continuum states
originating from rigid-wall boundary conditions~\cite{Scheuler2024}.
From this variation, one can obtain the density of states, shown in
Fig.~\ref{fig:stabilization}(b) and, finally, the linewidth of the
resonance can be extracted by a Lorentzian fit.
As a result, the partner state has the complex energy $E-i\Gamma/2$,
where the energy $E$ with respect to $E_{1,1}$ equals to
$201.2988\,$meV and $\Gamma = 0.1022\,$meV.
The BIC is lower by $2.8\,$meV and has $\Gamma=0$.
This gives an infinitely long nonradiative lifetime of the BIC.

The wavefunctions of the BIC and its partner state are presented in
Fig.~\ref{fig:3,606}.
The figures clearly show the difference between the partner state,
coupled to the continuum, and the BIC.
This can be seen by the oscillations that are visible for the partner
state along the $\rho$ axis.
These oscillations come from the resonant interference with the
continuum states above the threshold $E_{1,1}$.
However, the unambiguous missing of these oscillations for the BIC
as well as the exact exponential vanishing of its wavefunction with
increasing $\rho$ proves that we indeed found a bound state.
In this case, the interference is destructive resulting in an absence 
of nonradiative linewidth broadening.
One also observes that the oscillations in the partner state have a
wavelength of only about $10\,$nm.
For this reason we need a relatively large basis along the $\rho$ axis
to uncover these fine structures.
\begin{figure}
  \includegraphics[width = \columnwidth]{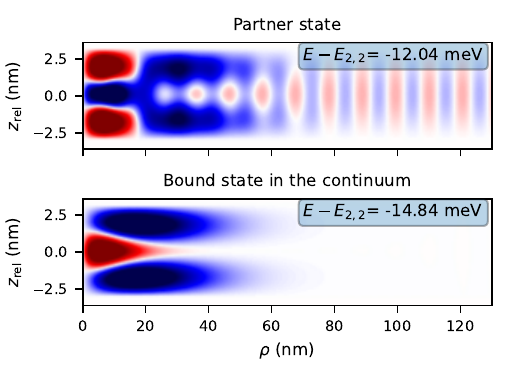}
  \caption{Wavefunctions of the BIC and its partner state sliced at
    $Z = 0\,$nm. The given energies of the states are with respect to
    the threshold $E_{2,2}$. The states were calculated for the QW width
    $L = 3.606\,$nm. The influence of the continuum can be seen by
    proper oscillations of the partner state with width $\Gamma =
    0.1022\,$meV along the $\rho$ axis, contrary to the BIC which does
    not couple to the continuum.}
\label{fig:3,606}
\end{figure}

%TK
%%%%%%%%%%%%%%%%%%%%%%%%%%%%%%%%%%%%%%%%%%%%%%%%%%%%%%%%%%%%%%%%%%%%%%%%%
\subsection{Oscillator strengths}
\begin{figure}
  \includegraphics[width = \columnwidth]{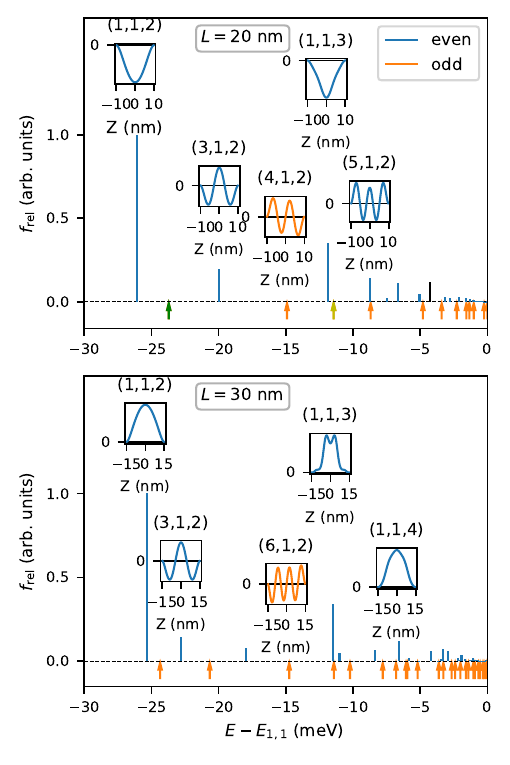}
  \caption{Oscillator strengths of exciton states in a cuprous oxide
    QW with the magnetic quantum number $m=1$. The upper and lower
    subfigures show results for QW width $L = 20\,$nm and $L =
    30\,$nm, respectively. Only states with even parity can have
    non-vanishing oscillator strengths. The orange arrows mark
    odd-parity states with vanishing oscillator strength. In the
    insets the derivatives of the wavefunctions~\eqref{eq:derivativeInset}
    are shown. The indicated quantum numbers of the states are
    $(n_{\mathrm{e}},n_{\mathrm{h}},n)$. The green and yellow arrows
    mark states discussed in Fig.~\ref{fig:zezh} and the black bar
    denotes the state shown in Fig.~\ref{fig:n5}.}
\label{fig:osci}
\end{figure}
Using Eq.~(\ref{eq:osci}) we estimate the transition rates between
the crystal ground state and the confined exciton states with angular
quantum number $m=1$.
The calculated relative oscillator strengths of the exciton
transitions in cuprous oxide QWs for the widths $L=20\,$nm and
$L=30\,$nm are shown in Fig.~\ref{fig:osci} by blue bars.
The maximal relative oscillator strength is normalized to unity.
In the insets the dependence of the derivative
\begin{equation}
  \lim_{\rho \to 0}  \frac{\partial}{\partial \rho} \psi(\rho,z_\mathrm{rel}=0, Z)
\label{eq:derivativeInset}
\end{equation}
upon $Z$ is shown for some states.
The underlying numerical approximations and extrapolations are discussed
in Sec.~\ref{subsec:oscillatorTheory} and Appendix.
The orange arrows mark odd-parity states with vanishing oscillator strengths.
There is a clear correlation between the nodal structures along the
confinement center-of-mass coordinate $Z$ in the insets and the
transition probabilities: states with an even number of peaks along
the $Z$ axis are characterized by a vanishing oscillator strength,
whereas states with an odd number of peaks are optically active.
This means that the confined exciton states of odd parity $\pi$ do not
produce light, i.e., they are dark states, whereas the even parity
states are bright confined excitons.
This observation highlights the fundamental role of parity in
determining the optical selection rules for excitonic transitions in QWs.
The wavefunctions discussed in Fig.~\ref{fig:zezh} have this odd
symmetry and therefore have vanishing oscillator strength.
These states are marked by the green and yellow arrow in the upper
panel in Fig.~\ref{fig:osci}.
The state (1,1,5) that is shown in Fig.~\ref{fig:n5} is marked by a
black bar in Fig.~\ref{fig:osci}.

Furthermore, a comparison between different QW widths reveals a systematic trend: 
as the QW width increases, the number of optically active
bound states also increases.
This behavior is consistent with the evolution of the excitonic energy
spectrum, where larger QW widths accommodate a greater number of
confined exciton states.
The observed enhancement in excitability with increasing $L$
underscores the interplay between quantum confinement and excitonic
optical properties, which is crucial for understanding and engineering
exciton-based optoelectronic devices.

%%%%%%%%%%%%%%%%%%%%%%%%%%%%%%%%%%%%%%%%%%%%%%%%%%%%%%%%%%%%%%%%%%%%%%%
\section{Conclusion}
\label{sec:conclusion}
In this article, we derived the relationships between the oscillator
strengths and the wavefunctions of Rydberg excitons in cuprous oxide QWs.
Due to the same parity of the conduction and valence bands in cuprous
oxide we had to analyze the symmetry properties of the next-to-leading
order terms in the momentum.
Moreover, the confinement along the $z$ axis makes the total wave
vector to be no longer a good quantum number, however, the wave vector
in the QW plane is still conserved.
As a result, the oscillator strength is related to the derivative of
the exciton wavefunction over the in-plane coordinate at the origin.

Using a parabolic approximation of the bands in the hydrogenlike
Schrödinger equation, we  calculated the wavefunctions of the exciton
states in cuprous oxide QW.
The calculated results include not only electron-hole bound states,
but also the electron-hole resonances associated to the quantum
confinement subbands of different symmetries.
The latter were obtained by the stabilization method.
Resonances appear above the lowest scattering threshold of a
particular parity and have nonzero linewidths.
However, among them there are BICs and their partner states that
result from the interference of resonances associated to different
subbands of the same symmetry.
The destructive interference brings the nonradiative broadening of
BICs towards zero, although the partner states are still broadened.
We calculated the wavefunctions of electron-hole bound states, a BIC
and a partner state.
The oscillations in the wavefunction of the partner state show its
resonant coupling to the continuum.
The absence of the oscillations of the BIC's wavefunction and its
exponential vanishing prove that we indeed obtained the BIC.

The calculated results allowed us to observe the symmetry properties
of the wavefunctions and their derivatives as well as their relations
to the oscillator strengths.
We obtained the relative values of the oscillator strengths for
exciton states of even parity, resulting from the corresponding
symmetries of the quantum confinement subbands.
The odd parity states, however, have vanishing oscillator strengths,
i.e., they correspond to the optically inactive exciton states.

The considered hydrogenlike excitons in cuprous oxide QWs result from
a two-band parabolic approximation of the cuprous oxide electronic
energy structure near the $\Gamma$ point~\cite{French2009}.
The simplest parabolic valence band represents only the dominant
diagonal part of the Suzuki-Hensel Hamiltonian~\cite{Suzuki1974,Schweiner17b}.
It thus disregards the effects of the valence band on the yellow
exciton series in bulk Cu$_{2}$O studied, for example, in
Refs.~\cite{Kaz14,Schoene16,Schweiner17b}.
In narrow QWs and for exciton Rydberg states, these perturbations of
the band structure are obscured by the strong quantum confinement
effect~\cite{Belov2024}.
In our work, we employed infinite QW potentials and constant
dielectric permittivity leading to pure Coulomb interactions in a
finite domain.
More precise consideration requires finite QW potentials and different
dielectric constants in the QW and in the barriers.
The latter has been studied in Refs.~\cite{Rytova,Keldysh} and leads to
a distortion of the Coulomb interaction at short length scales.
The effect of this modification on the energy spectra of excitons in
Cu$_2$O QWs has been discussed in Ref.~\cite{Belov2024}.
The mentioned finer effects can additionally distort the wavefunction
and, as a result, smear the exact symmetry-induced optical properties
of the exciton in cuprous oxide QW.
The effects of the full complex valence band structure require a
dedicated analysis, which is currently in progress.

%%%%%%%%%%%%%%%%%%%%%%%%%%%%%%%%%%%%%%%%%%%%%%%%%%%%%%%%%%%%%%%%%%

\acknowledgments
This work was supported by Deutsche Forschungsgemeinschaft (DFG) through the
DFG Priority Programme 1929 ``Giant interactions in Rydberg systems'' (GiRyd),
Grants No.\ MA 1639/16-1 and No.\ SCHE 612/4-2.

%%%%%%%%%%%%%%%%%%%%%%%%%%%%%%%%%%%%%%%%%%%%%%%%%%%
\appendix*
\section{Extrapolation of functions}
%\label{sec:appendix}
%
\begin{figure}[b]
  \includegraphics[width = \linewidth]{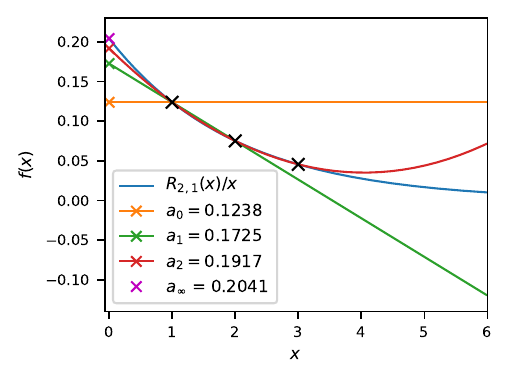}
  \caption{Extrapolation of the hydrogenic wavefunction $R_{2,1}(x)$
    shown in blue. The approximations of this function by polynomials
    of different orders (from zero to two) are shown by yellow, green,
    and red colors. The equidistant points which are used for the
    approximation are indicated by black crosses. The extrapolated
    value $a_{\infty}$ is shown by the magenta cross.}
\label{fig:explanation}
\end{figure}
Here we explain the numerical evaluation of the derivative of the
wavefunction in the limit $\rho\to 0$.
As $\psi(\rho = 0) = 0$, the derivative in $\rho$ direction can, in
the limit $\rho \to 0$, be replaced by $\psi(\rho)/\rho$.
However, this is numerically tricky to calculate. 
The solution to this problem is an approximation of the wavefunction,
using a simple polynomial model.
The extrapolation of $n+1$ equidistant data points $f_i$ by a
degree-$n$ polynomial $P_{n}(x)$ to $x=0$ is given by the formula
\begin{align}
  a_n  \equiv P_{n}(x = 0) = \sum_{i=1}^{n+1} (-1)^{i+1} \binom{n+1}{i} f_i \,.
\label{eq:extrapolation}
\end{align}
This is illustrated in Fig.~\ref{fig:explanation} for the radial
wavefunction $R_{2,1}(x)$ of the hydrogen atom.
By assuming an exponential decrease in the error of the approximations
$a_n$, these values can be extrapolated to $n\to\infty$, i.e.
\begin{align}
  a_{\infty} = \frac{a_{n}^2-a_{n-1}a_{n+1}}{2a_{n} - a_{n-1} - a_{n+1}} \, .
\end{align}
For the example in Fig.~\ref{fig:explanation} the value of
$a_{\infty}$ is accurate to about eight decimal places.

%\bibliography{paper}

\end{document}